\documentclass[11pt]{article}

\usepackage[T1]{fontenc}
\usepackage[utf8]{inputenc}
\usepackage{lmodern}
\usepackage[a4paper,margin=2.5cm]{geometry}
\usepackage{authblk}
\usepackage{amsmath,amssymb}
\usepackage{graphicx}
\usepackage{booktabs}
\usepackage{microtype}
\usepackage{setspace}
\usepackage{lineno}
\usepackage[numbers,sort&compress]{natbib}
\usepackage[hidelinks]{hyperref}
\usepackage{xurl}
\usepackage{caption}

\graphicspath{{./}}
\newcommand{\ket}[1]{\lvert #1\rangle}
\newcommand{\bra}[1]{\langle #1\rvert}

\title{Non-Markovian and non-Condon vibrationally assisted electron transfer in an effective ligand--receptor complex}

\author[1]{Muhammad Waqas Haseeb}
\author[2]{Mohamad Toutounji\thanks{Correspondence: mtotounji@uaeu.ac.ae}}
\affil[1]{Department of Physics, United Arab Emirates University, Al Ain, United Arab Emirates}
\affil[2]{Department of Chemistry, United Arab Emirates University, Al Ain, United Arab Emirates}
\date{}

\begin{document}

\maketitle

\begin{abstract}
We investigate vibrationally assisted electron transfer in an effective ligand--receptor complex using finite-temperature non-Markovian quantum-state diffusion with explicitly colored environmental noise. The model comprises donor and acceptor electronic states, an explicit bound-ligand vibrational mode, and a Gaussian protein--solvent bath. Linearization of a coordinate-dependent diabatic Hamiltonian yields two distinct system--environment interactions: diagonal coupling, which modulates the donor--acceptor energy gap, and off-diagonal non-Condon coupling, which modulates the electronic tunneling matrix element. In the rapidly equilibrating diagonal-bath limit, the model recovers the expected Marcus--Levich--Jortner vibronic-transfer behavior. Persistent environmental correlations instead generate non-exponential dynamics characterized by transient oscillations, back-transfer, and memory-dependent damping. Under matched weak-tunneling conditions, off-diagonal coupling opens a direct transition pathway and produces a larger finite-time acceptor population throughout the investigated parameter space. Because these dynamics do not generally support a unique kinetic rate, the enhancement is quantified using a time-averaged transfer yield. Comparisons with exact closed-system and commuting pure-dephasing limits validate the deterministic propagator and colored-noise construction, respectively, while the noncommuting dynamics are evaluated within a declared first-order finite-memory closure. These results identify ligand vibration, environmental memory, and non-Condon tunneling modulation as cooperating mechanisms that can dynamically regulate electron transfer in receptor-like environments, while maintaining a clear distinction between a mechanistic coarse-grained model and an atomistically parameterized biological complex.

\end{abstract}

\section*{Why it matters}

Electron transfer at a ligand--receptor interface can be controlled in two physically different ways: nuclear motion can change the energetic gap between charge-localized states, or it can change the electronic overlap that permits tunneling. A structured protein--solvent environment can remember earlier motion long enough to reinforce, redirect, or damp these pathways. This work derives both coupling mechanisms in one ligand--receptor model and shows that off-diagonal, non-Condon fluctuations can act as a tunneling gate. It also distinguishes a true kinetic rate from enhanced short-time transferred population, which is essential when environmental memory produces non-exponential dynamics.

\textbf{Keywords:} ligand--receptor interaction; electron transfer; vibrational gating; non-Condon coupling; non-Markovian dynamics; NMQSD

\section{Introduction}

Electron transfer (ET) is a basic step in respiration, photosynthesis, redox catalysis, enzymatic turnover, and molecular recognition. In biological systems the electron commonly moves between localized donor and acceptor configurations through a protein, solvent, or molecular bridge. Its probability is therefore controlled not only by the donor--acceptor energy difference, but also by nuclear reorganization and by the electronic coupling of the two states \cite{de1966studies,MarcusSutin1985,gray2003electron,moser2010guidelines,haseeb2025non,Haseeb2024vibration}. Marcus theory organizes the classical contribution of the environment, while Marcus--Levich--Jortner (MLJ) theory resolves the participation of a quantized intramolecular vibration into vibronic channels \cite{Jortner1976,heller2020semiclassical,may2023charge,haseeb2026non,Haseeb2026nonmarkovian}. These theories provide the natural kinetic reference when electronic tunneling is weak, the nuclear coordinates equilibrate rapidly, and the populations relax approximately exponentially.

A bound ligand can participate in this process through its internal vibrations or through collective motion at the binding interface. Vibrationally assisted electron transfer has consequently been discussed as a possible molecular-recognition mechanism, most prominently in models of phonon-assisted tunneling in olfaction \cite{Turin1996,brookes2007could,solov2012vibrationally,bittner2012quantum,a92}. The broader physical idea does not depend on a particular receptor: a ligand displacement can bring an electronic energy gap into vibronic resonance, alter a bridge geometry, or change donor--acceptor orbital overlap. In this sense, the vibration acts as a dynamical coordinate of the ligand--receptor interaction rather than merely as a passive heat bath.

Two distinct coupling mechanisms must then be separated. In an energy-gap or diagonal interaction, nuclear motion changes the donor and acceptor energies while the electronic tunneling element is held fixed. This is the usual Condon-like spin--boson geometry. In a non-Condon interaction, the tunneling matrix element itself depends on nuclear position. Protein motion, ligand orientation, side-chain rearrangement, and interfacial electrostatics can all modulate the distance, barrier, or orbital overlap along an ET pathway. The resulting off-diagonal fluctuations can enhance or suppress transfer even when energy-gap matching is unfavorable \cite{Skourtis2005NonCondon,Milischuk2007,Skourtis2010FluctuationsET,Acharyya2020NonDiagonalET,a2,a3,a4}. A central question is therefore whether ligand-induced nuclear motion assists ET by changing the activation energetics, by opening the tunneling pathway, or by a combination of the two.

The temporal structure of the environment is equally important. Protein, membrane, solvent, and intramolecular coordinates span many timescales. If their correlations decay much faster than the electronic and vibrational motion, their action can be approximated as Markovian noise. If the correlation time is comparable to the transfer dynamics, the environment retains memory. The system at time $t$ then responds to its earlier states through a retarded kernel, and population evolution can become oscillatory, multi-exponential, or dependent on the phase of the bath response \cite{Breuer2002theory,Tanimura2020,a5,a6}. Environmental memory is not intrinsically beneficial: depending on the coupling operator and timescale, it can open a transient channel, return amplitude to the donor, or damp coherent exchange.

Non-Markovian quantum-state diffusion (NMQSD) provides a trajectory representation in which the bath spectral density determines both colored fluctuations and retarded feedback \cite{diosi1997non,Strunz1999,strunz1999open,a11}. The formal NMQSD equation is exact for a Gaussian bath before its functional derivative is closed. Practical calculations require a closure, hierarchy, or path-integral construction, and the approximation actually propagated must therefore be stated separately from the exact representation \cite{Hartmann2017exact,Hsieh2018I,Makri1998,Tanimura2020,a7,a8,a9,a10}. Here we use a normalized first-order finite-memory closure together with explicitly colored quantum and thermal noise. Its domain is assessed through covariance, time-step, basis, and limiting-case tests reported in the Supporting Information (SI).

Our aim is to determine how environmental memory and non-Condon coupling regulate vibrationally assisted ET in a receptor-like setting. We represent the receptor ET unit by donor and acceptor diabatic states, the ligand by one effective vibrational coordinate, and the surrounding protein--solvent matrix by either a rapidly decaying or a structured correlation function. We first connect the model to MLJ theory in a controlled kinetic limit. We then compare diagonal and off-diagonal coupling under matched conditions and follow their time-resolved donor and acceptor populations. When the dynamics do not support a unique rate, we use a finite-time transfer yield that measures acceptor occupation without imposing exponential kinetics.

The model is intentionally coarse grained. It describes a class of ligand--receptor interactions and is not an atomistic simulation of a named complex. Thus, the results establish mechanisms and parameter dependencies, not a quantitative rate for a particular receptor. A system-specific application would require electronic-structure and molecular-dynamics calculations to identify the diabatic states, ligand modes, electronic-coupling derivatives, and diagonal, off-diagonal, and cross-correlated spectral densities. This boundary is retained throughout the biological interpretation.

\section{Theory and methods}

\subsection{Effective ligand--receptor Hamiltonian}

We use localized receptor electronic states $\{\ket{D},\ket{A}\}$, which may represent redox centers or two charge-localized configurations connected through the ligand--receptor interface. For nuclear coordinates $\mathbf q$, the real two-state diabatic Hamiltonian is
\begin{equation}
H_{\mathrm{el}}(\mathbf q)=
\begin{pmatrix}
E_D(\mathbf q) & V_{DA}(\mathbf q)\\
V_{DA}(\mathbf q) & E_A(\mathbf q)
\end{pmatrix}.
\label{eq:diabatic-matrix}
\end{equation}
Writing $\delta E(\mathbf q)=E_D(\mathbf q)-E_A(\mathbf q)$ and removing the state-independent mean energy gives
\begin{equation}
H_{\mathrm{el}}(\mathbf q)=
\frac{\delta E(\mathbf q)}{2}\sigma_z+V_{DA}(\mathbf q)\sigma_x.
\label{eq:pauli-form}
\end{equation}
One effective ligand vibration, with $Q_v=b+b^\dagger$, is retained inside the system. Linearizing its contribution to the energy gap at the reference binding geometry gives $\delta E(Q_v)=\epsilon+2\gamma Q_v$. With $V_{DA}^{(0)}=\Delta/2$ and $\hbar=1$ during propagation, the system Hamiltonian becomes
\begin{equation}
H_{\mathrm S}=\frac{\epsilon}{2}\sigma_z+\frac{\Delta}{2}\sigma_x
+\omega_v b^\dagger b+\gamma\sigma_z(b+b^\dagger).
\label{eq:system-hamiltonian}
\end{equation}
Here $\epsilon$ is the effective reaction-energy bias, $\Delta$ is twice the bare donor--acceptor coupling, $\omega_v$ is the ligand-mode frequency, and $\gamma$ measures diagonal vibronic coupling. Completing the square on each electronic surface gives the displacement $d=\gamma/\omega_v$, the Huang--Rhys factor, and the corresponding mode reorganization energy,
\begin{equation}
S=(2d)^2=\left(\frac{2\gamma}{\omega_v}\right)^2,
\qquad
\lambda_v=S\omega_v=\frac{4\gamma^2}{\omega_v}.
\label{eq:huang-rhys}
\end{equation}
The donor is initialized with its equilibrated displaced oscillator state, $\ket{\Psi(0)}=\ket D\otimes\ket{-d}$. The detailed displacement convention is derived in Note~S2.

\subsection{Derivation of diagonal and off-diagonal environmental coupling}

The remaining ligand, receptor, protein, and solvent coordinates are represented by harmonic modes $q_k\propto B_k+B_k^\dagger$. Expanding both parts of Eq.~\eqref{eq:pauli-form} to first order gives
\begin{align}
\delta E(\mathbf q)&\simeq \epsilon+2\gamma Q_v+\sum_k c_{zk}(B_k+B_k^\dagger),\\
2V_{DA}(\mathbf q)&\simeq \Delta+\sum_k c_{xk}(B_k+B_k^\dagger).
\label{eq:coordinate-expansion}
\end{align}
Substitution into Eq.~\eqref{eq:pauli-form} yields the generalized interaction
\begin{equation}
H_{\mathrm{SB}}=L_z\otimes X_z+L_x\otimes X_x,
\qquad
L_z=\frac{\sigma_z}{2},\quad L_x=\frac{\sigma_x}{2},
\label{eq:coupling-operators}
\end{equation}
where $X_\alpha=\sum_kc_{\alpha k}(B_k+B_k^\dagger)$. The $L_z$ term changes the instantaneous energy gap and therefore represents diagonal environmental coupling. The $L_x$ term changes $V_{DA}$ itself and is the leading non-Condon correction to a coordinate-independent tunneling element. The factor of one half follows directly from the definitions of the gap and $2V_{DA}$ in Eq.~\eqref{eq:coordinate-expansion}; the coupling magnitude is contained in $X_\alpha$ and its spectral density.

In a molecular complex, $X_z$ and $X_x$ may coexist and be cross-correlated. To isolate their physical roles, the present calculations use one operator at a time under matched bath conditions:
\begin{equation}
H_{\mathrm{tot}}^{(\alpha)}=H_{\mathrm S}+\sum_k\omega_kB_k^\dagger B_k
+L_\alpha\otimes X_\alpha,
\qquad \alpha\in\{z,x\}.
\label{eq:total-hamiltonian}
\end{equation}
The two cases are therefore limiting coupling geometries of the same ligand--receptor Hamiltonian, not two differently parameterized biological systems.

\subsection{Environmental spectra and memory}

For either coupling geometry, the bath is specified by
\begin{equation}
J_\alpha(\omega)=\pi\sum_kc_{\alpha k}^2\delta(\omega-\omega_k),
\qquad
\lambda_\alpha=\frac{1}{\pi}\int_0^\infty\frac{J_\alpha(\omega)}{\omega}\,d\omega.
\label{eq:spectral-definition}
\end{equation}
We compare a broad Ohmic spectrum and an underdamped Brownian-oscillator spectrum,
\begin{align}
J_{\mathrm{Ohm}}(\omega)&=\pi\frac{\lambda_s}{\omega_c}\omega e^{-\omega/\omega_c},\\
J_{\mathrm{BO}}(\omega)&=
\frac{2\lambda_{\mathrm{BO}}\Gamma\Omega_0^2\omega}
{(\Omega_0^2-\omega^2)^2+\Gamma^2\omega^2}.
\label{eq:spectral-models}
\end{align}
The associated finite-temperature correlation function is
\begin{equation}
C_\alpha(t)=\frac{1}{\pi}\int_0^\infty d\omega\,J_\alpha(\omega)
\left[\coth\!\left(\frac{\omega}{2k_{\mathrm B}T}\right)\cos(\omega t)
-i\sin(\omega t)\right].
\label{eq:correlation}
\end{equation}
The Ohmic cutoff $\omega_c$ sets a short correlation time, whereas $\Omega_0$ and $\Gamma$ set the oscillation and envelope times of the structured bath. Markovianity is therefore assessed by comparing the decay of $C(t)$ with the electronic and vibrational timescales; it is not inferred from the name of the spectrum alone.

\subsection{NMQSD equation and finite-memory operators}

After tracing out a Gaussian environment, the linear NMQSD representation is
\begin{equation}
\frac{d}{dt}\ket{\psi_t(z^*)}=
\left[-iH_{\mathrm S}+Lz_t^*
-L^\dagger\int_0^t ds\,C(t-s)\frac{\delta}{\delta z_s^*}\right]
\ket{\psi_t(z^*)},
\label{eq:nmqsd}
\end{equation}
with
\begin{equation}
\mathbb E[z_t]=\mathbb E[z_tz_s]=0,
\qquad \mathbb E[z_tz_s^*]=C(t-s).
\label{eq:noise-covariance}
\end{equation}
Equation~\eqref{eq:nmqsd} is formally exact, but the functional derivative prevents direct propagation. Introducing $\delta\ket{\psi_t}/\delta z_s^*=O(t,s,z^*)\ket{\psi_t}$ and expanding over the memory interval $\tau=t-s$ gives the first-order closure used here,
\begin{equation}
O(t,s,z^*)\simeq L-i\tau[H_{\mathrm S},L],
\qquad
\overline O^{(1)}(t)=g_0(t)L-i g_1(t)[H_{\mathrm S},L],
\label{eq:o-expansion}
\end{equation}
where, in the finite-temperature decomposition used for propagation,
\begin{equation}
g_n(t)=\int_0^t d\tau\,\tau^n C_0(\tau),
\qquad
C_0(t)=\frac{1}{\pi}\int_0^\infty J(\omega)e^{-i\omega t}\,d\omega.
\label{eq:memory-coefficients}
\end{equation}
The thermal part of Eq.~\eqref{eq:correlation} is included as an independent real colored potential; the complete normalized propagation equation is given in Note~S4.

The operator dependence of the memory correction follows directly from the Pauli algebra. For diagonal coupling,
\begin{equation}
[H_{\mathrm S},L_z]=-i\frac{\Delta}{2}\sigma_y,
\qquad
\overline O_z^{(1)}(t)=\frac{g_0(t)}{2}\sigma_z
-\frac{\Delta g_1(t)}{2}\sigma_y.
\label{eq:obar-z}
\end{equation}
Thus, a diagonal bath changes population only in cooperation with the bare tunneling term. For off-diagonal coupling,
\begin{equation}
[H_{\mathrm S},L_x]=i\left[\frac{\epsilon}{2}\sigma_y
+\gamma\sigma_y(b+b^\dagger)\right],
\label{eq:commutator-x}
\end{equation}
and hence
\begin{equation}
\overline O_x^{(1)}(t)=\frac{g_0(t)}{2}\sigma_x
+g_1(t)\left[\frac{\epsilon}{2}\sigma_y
+\gamma\sigma_y(b+b^\dagger)\right].
\label{eq:obar-x}
\end{equation}
Equation~\eqref{eq:obar-x} displays the non-Condon mechanism explicitly: the bath acts along the tunneling coordinate, while its memory correction mixes that coordinate with both the electronic bias and the ligand vibration. In the short-memory limit $g_1(t)$ becomes negligible and the feedback is local. When $g_1(t)$ persists, earlier bath fluctuations feed back into the transfer amplitude and can produce non-exponential exchange. Equations~\eqref{eq:obar-z} and \eqref{eq:obar-x} are the diagonal and off-diagonal operators implemented in the simulations.

\subsection{Rates, populations, and finite-time transfer}

The ensemble density operator and reduced electronic populations are
\begin{equation}
\rho_{\mathrm{full}}(t)=\mathbb E[\ket{\widetilde\psi_t}\bra{\widetilde\psi_t}],
\quad
\rho_{\mathrm{el}}(t)=\operatorname{Tr}_{\mathrm{vib}}\rho_{\mathrm{full}}(t),
\quad
P_{D/A}(t)=\rho_{DD/AA}(t).
\label{eq:populations}
\end{equation}
In the weak-tunneling classical-solvent limit, the reference vibronic rate is
\begin{equation}
k_{\mathrm{MLJ}}(\epsilon)=\frac{2\pi}{\hbar}|V|^2
\frac{1}{\sqrt{4\pi\lambda_s k_{\mathrm B}T}}
\sum_{m=0}^\infty e^{-S}\frac{S^m}{m!}
\exp\!\left[-\frac{(\epsilon-\lambda_s-m\omega_v)^2}
{4\lambda_s k_{\mathrm B}T}\right],
\qquad V=\frac{\Delta}{2}.
\label{eq:mlj}
\end{equation}
The channel centers $\epsilon_m=\lambda_s+m\omega_v$ express vibronic energy matching, while $\gamma$ changes their Franck--Condon weights through $S$.

A scalar ET rate is reported only if the population satisfies the kinetic tests in Table~S2. Memory-bearing off-diagonal trajectories frequently fail these tests because they are oscillatory or multi-exponential. For the matched coupling-geometry comparison we therefore use
\begin{equation}
Y_\tau^{(\alpha)}=\frac{1}{\tau}\int_0^\tau P_A^{(\alpha)}(t)\,dt,
\qquad
\Delta Y_\tau=Y_\tau^{(x)}-Y_\tau^{(z)}.
\label{eq:yield}
\end{equation}
$Y_\tau$ is a bounded measure of population delivered to the acceptor during a specified observation time and does not assume irreversible exponential transfer.

\subsection{Numerical implementation}

Stationary colored processes are generated by frequency-domain synthesis from the target spectral weights, and antithetic trajectories reduce odd-moment sampling error. The system Hamiltonian is propagated by a second-order split scheme; the nonlinear bath terms are advanced at the midpoint and each trajectory is normalized. All physical and numerical parameters, trajectory counts, basis sizes, fitting windows, covariance errors, and convergence checks are reported in Tables~S1--S3 and Figs.~S1--S6. 

\section{Results}

\subsection{Vibronic energy matching establishes the diagonal-coupling reference}

\begin{figure}[p]
\centering
\includegraphics[width=\linewidth]{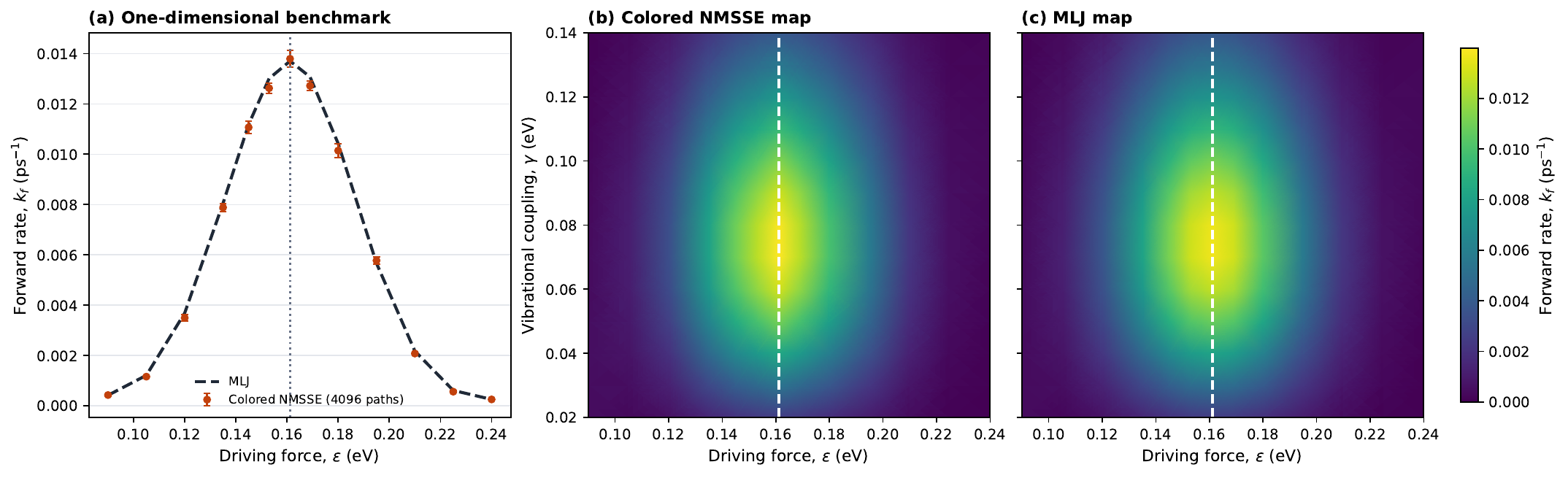}
\caption{\textbf{Vibronic-transfer reference for diagonal coupling.} (a) Forward ET rate from colored classical-solvent propagation and MLJ theory. (b,c) Corresponding rate maps as the electronic bias and ligand-mode coupling are varied. The vertical line identifies the one-quantum vibronic channel. Parameters and numerical diagnostics are listed in Tables~S1 and S3.}
\label{fig:1}
\end{figure}

We first place the ligand--receptor model in the weak-tunneling, rapidly equilibrating classical-solvent regime, where the population dynamics admit a kinetic description. Figure~\ref{fig:1}(a) presents a one-dimensional comparison between the forward rates obtained from the colored stochastic propagation and the Marcus--Levich--Jortner expression in Eq.~\eqref{eq:mlj}. The stochastic results, including their sampling uncertainties, follow the MLJ curve across the investigated driving-force range. Both approaches attain their maximum at the one-quantum activationless condition,
\begin{equation}
\epsilon=\lambda_s+\omega_v,
\end{equation}
where the electronic energy released during transfer compensates the solvent reorganization energy and creates one quantum of the ligand vibration. Moving away from this condition reduces the rate because an additional solvent activation penalty is required.

Figure~\ref{fig:1}(b) extends the colored stochastic calculation over both the driving force $\epsilon$ and the ligand-mode coupling $\gamma$. The resulting rate map contains a vertical vibronic ridge centered at the same one-quantum resonance. The position of this ridge is determined by energy matching and therefore remains approximately independent of $\gamma$. In contrast, its intensity varies non-monotonically with the ligand displacement because $\gamma$ controls the Huang--Rhys factor,
\begin{equation}
S=\left(\frac{2\gamma}{\omega_v}\right)^2,
\end{equation}
and hence the one-quantum Franck--Condon weight $e^{-S}S$. At weak displacement the overlap with the one-quantum acceptor state is small, whereas at larger displacement the vibronic intensity is progressively redistributed into higher-order channels. The strongest one-quantum transfer therefore occurs at an intermediate ligand-mode coupling.

Figure~\ref{fig:1}(c) shows the corresponding MLJ rate map calculated on the same parameter grid and displayed using the same color scale. Its resonance position, width, and dependence on $\gamma$ closely reproduce the colored stochastic map in Fig.~\ref{fig:1}(b). The agreement among the one-dimensional comparison and the two rate maps demonstrates that diagonal energy-gap fluctuations recover the conventional reorganization-controlled mechanism in the kinetic limit. Figure~\ref{fig:1} therefore establishes the reference behavior against which the non-Markovian memory effects and off-diagonal non-Condon tunneling mechanism are examined below. Quantitative agreement and convergence diagnostics are provided in Note~S5 and Table~S3.

\subsection{The bath spectrum sets the environmental memory seen by the complex}

\begin{figure}[p]
\centering
\includegraphics[width=\linewidth]{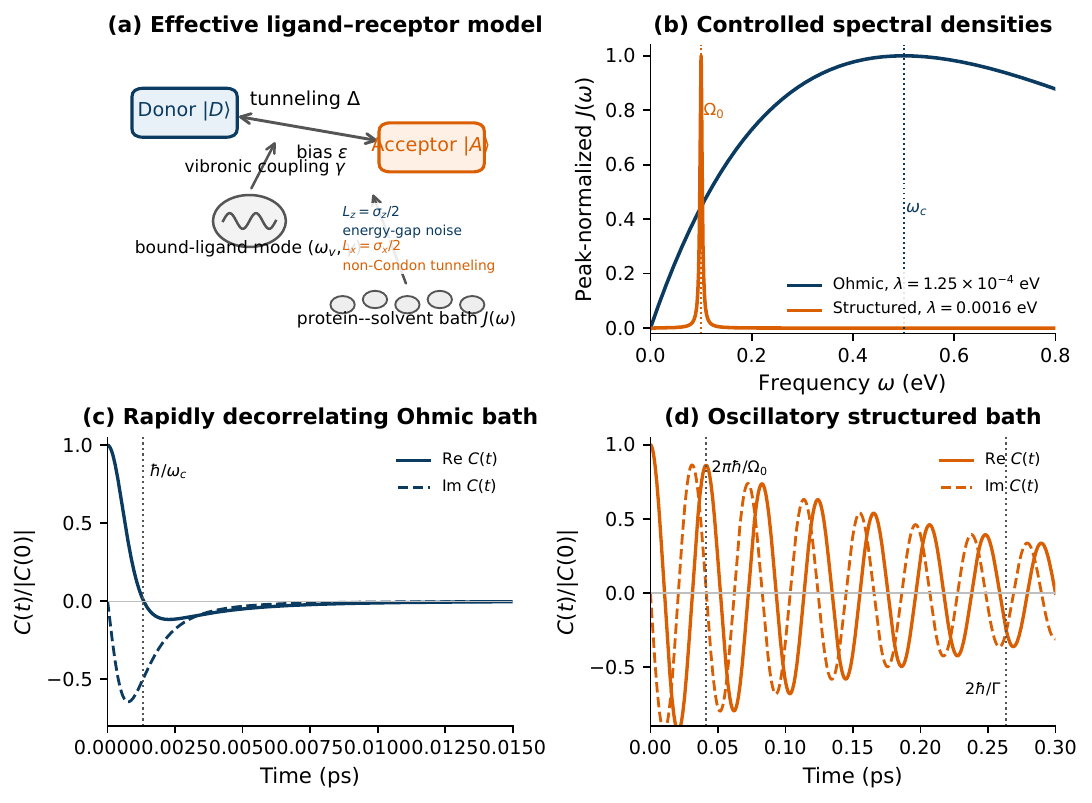}
\caption{\textbf{Effective ligand--receptor model and environmental memory.} (a) Donor and acceptor states, the explicit ligand vibration, and diagonal ($L_z$) or off-diagonal ($L_x$) coupling to the residual protein--solvent bath. (b) Ohmic and structured spectral shapes. (c,d) Their normalized correlation functions, showing short and oscillatory memory, respectively. Parameters are listed in Table~S1.}
\label{fig:2}
\end{figure}

Figure~\ref{fig:2} defines how the effective ligand--receptor complex is coupled to its surrounding protein--solvent environment. In Fig.~\ref{fig:2}(a), the donor and acceptor states are separated by the electronic bias $\epsilon$ and connected by the bare tunneling element $\Delta$. A bound-ligand vibration of frequency $\omega_v$ is retained explicitly and coupled to the electronic energy gap with strength $\gamma$. The remaining nuclear coordinates form a residual bath described by the spectral density $J(\omega)$. These coordinates can act through the diagonal operator $L_z=\sigma_z/2$, which shifts the donor and acceptor energies in opposite directions, or through the off-diagonal operator $L_x=\sigma_x/2$, which modulates the donor--acceptor tunneling matrix element. The latter interaction is the non-Condon pathway because nuclear motion directly changes the electronic coupling.

Figure~\ref{fig:2}(b) compares the two spectral shapes used to represent the residual environment. The broad Ohmic spectrum distributes its weight over a wide frequency range, with the cutoff frequency $\omega_c$ setting the dominant response timescale. The structured spectrum is concentrated around the characteristic frequency $\Omega_0$ and represents a collective protein--solvent or ligand--receptor motion with a relatively well-defined frequency. Both spectra are peak normalized in this panel so that their shapes and frequency scales can be compared directly; their plotted peak heights therefore do not represent equal reorganization energies. The Ohmic bath provides a rapidly varying background, whereas the narrow structured spectrum is capable of producing a longer-lived, frequency-selective environmental response.

The corresponding Ohmic correlation function obtained from Eq.~\eqref{eq:correlation} is shown in Fig.~\ref{fig:2}(c). Its real part describes the temporal persistence of the environmental fluctuations, while its imaginary part represents the associated reactive bath response. Both components decay rapidly on the characteristic timescale
\begin{equation}
\tau_{\mathrm{Ohm}}\sim\frac{\hbar}{\omega_c}.
\end{equation}
Consequently, the bath loses information about its earlier interaction with the donor--acceptor subsystem before substantial electronic or ligand-vibrational motion occurs. On the relevant system timescale, this environment therefore approaches a short-memory or effectively local fluctuating limit.

In contrast, Fig.~\ref{fig:2}(d) shows that the structured bath retains pronounced oscillations in both the real and imaginary components of $C(t)$. The recurrence period and the decay time of the oscillatory envelope are approximately
\begin{equation}
\tau_{\mathrm{osc}}=\frac{2\pi\hbar}{\Omega_0},
\qquad
\tau_{\mathrm{env}}\simeq\frac{2\hbar}{\Gamma},
\end{equation}
respectively. Because the correlation remains finite over several oscillation cycles, the environment can return amplitude and phase information to the ligand--receptor subsystem rather than acting only as instantaneous noise. Depending on whether this returning fluctuation acts through $L_z$ or $L_x$, it can reinforce transfer, promote back-transfer, or alter the damping and phase of the donor--acceptor population exchange. The short-lived correlation in Fig.~\ref{fig:2}(c) and the oscillatory memory in Fig.~\ref{fig:2}(d) therefore provide the physical basis for the contrasting population dynamics examined in Figs.~\ref{fig:3} and \ref{fig:4}.

\subsection{Non-Markovian memory reshapes electron-transfer dynamics}

\begin{figure}[p]
\centering
\includegraphics[width=\linewidth]{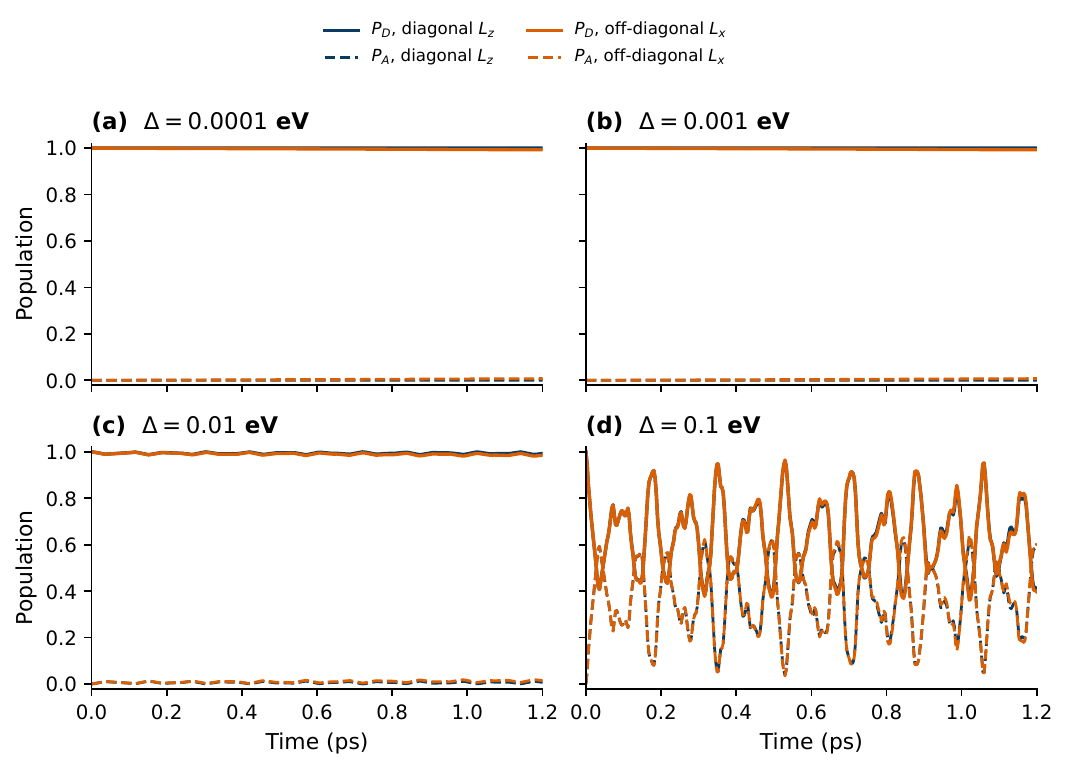}
\caption{\textbf{Ligand--receptor population dynamics in a short-memory Ohmic environment.} Donor and acceptor populations are compared for diagonal and off-diagonal coupling as the bare electronic tunneling is increased from weak to strong. Shading denotes trajectory-block uncertainty. Complete parameters are provided in Table~S1.}
\label{fig:3}
\end{figure}

Figure~\ref{fig:3} shows the donor and acceptor populations in the rapidly decorrelating Ohmic environment as the bare electronic tunneling element $\Delta$ is increased. Solid and dashed curves represent the donor and acceptor populations, respectively, while the diagonal $L_z$ and off-diagonal $L_x$ coupling geometries are shown separately. Because the Ohmic correlation decays much faster than the electronic and ligand-vibrational dynamics, an environmental fluctuation acts only briefly before losing its phase relation with the ligand--receptor subsystem.

At the smallest tunneling element, Fig.~\ref{fig:3}(a), diagonal coupling leaves the electron almost completely localized on the donor. The operator $L_z=\sigma_z/2$ changes the instantaneous donor--acceptor energy gap but does not directly connect the two electronic states. Population transfer must therefore be initiated by the very small Hamiltonian tunneling term. Off-diagonal coupling produces a larger, although still weak, acceptor population because $L_x=\sigma_x/2$ acts directly along the electronic-transition coordinate. The short-lived bath fluctuations, however, cannot maintain this tunneling gate for a sufficiently long time to generate substantial transfer.

When $\Delta$ is increased in Fig.~\ref{fig:3}(b), the diagonal mechanism begins to acquire a small transfer contribution through the bare electronic coupling. The off-diagonal bath still provides the more effective pathway because its fluctuations directly modulate the donor--acceptor matrix element. Nevertheless, both acceptor populations remain small, showing that rapid environmental decorrelation prevents the repeated or phase-coordinated action required for a pronounced non-Condon gating response.

At the intermediate tunneling strength in Fig.~\ref{fig:3}(c), Hamiltonian mixing becomes comparable to the weak bath-induced pathway. Both coupling geometries now produce visible donor--acceptor population exchange, although the off-diagonal interaction continues to yield a somewhat larger acceptor population. The difference is modest because each Ohmic fluctuation decays before it can feed coherently back into the subsequent electronic motion. The bath consequently acts mainly as a source of short-lived modulation and damping rather than as a persistent dynamical coordinate.

\begin{figure}[p]
\centering
\includegraphics[width=\linewidth]{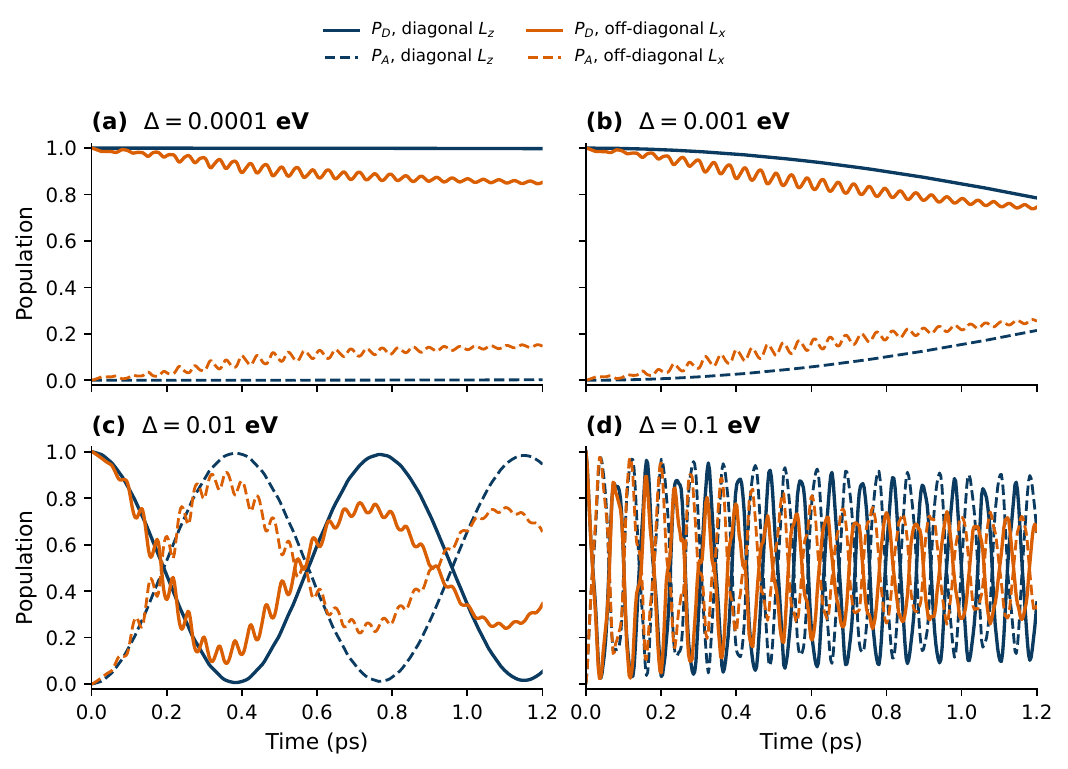}
\caption{\textbf{Ligand--receptor population dynamics in a structured non-Markovian environment.} The same coupling geometries and tunneling progression as in Fig.~\ref{fig:3} are shown for a bath with an oscillatory memory kernel. Off-diagonal coupling opens a pronounced weak-tunneling pathway, whereas at stronger tunneling memory mainly reshapes coherent exchange. Parameters are listed in Table~S1.}
\label{fig:4}
\end{figure}

In the strong-tunneling regime of Fig.~\ref{fig:3}(d), the donor and acceptor populations undergo large-amplitude oscillatory exchange. The nearly overlapping $L_z$ and $L_x$ curves demonstrate that the population dynamics are now governed primarily by the system Hamiltonian and the explicit ligand vibration. Changing the bath-coupling operator produces only small differences in the oscillation phase and damping because the rapidly decorrelating environment cannot compete with the intrinsic tunneling motion. Figure~\ref{fig:3} therefore shows that a fast bath can open a weak off-diagonal pathway when the bare tunneling is small, but it does not sustain a distinct memory-controlled mechanism. This behavior provides the short-memory reference for the structured non-Markovian dynamics considered in Fig.~\ref{fig:4}.

Figure~\ref{fig:4} shows the donor and acceptor populations for the structured environment as the bare tunneling element $\Delta$ is increased. Solid and dashed curves denote the donor and acceptor populations, respectively, while the diagonal $L_z$ and off-diagonal $L_x$ interactions are compared under the same structured-bath conditions. Unlike the rapidly decorrelating environment in Fig.~\ref{fig:3}, the structured correlation persists for several oscillation cycles. A fluctuation can therefore return to the ligand--receptor subsystem with a non-random phase and modify the subsequent transfer dynamics.

At the smallest tunneling element, $\Delta=10^{-4}\,\mathrm{eV}$, Fig.~\ref{fig:4}(a) displays the clearest distinction between the two coupling geometries. Diagonal coupling leaves the electron almost completely localized on the donor because $L_z$ changes only the donor--acceptor energy gap and the bare Hamiltonian provides essentially no electronic mixing. In contrast, the off-diagonal interaction generates a substantial acceptor population accompanied by small oscillatory modulations. Here $L_x$ directly connects the donor and acceptor states, while the structured bath repeatedly modulates this transition coordinate. The gradual accumulation of acceptor population therefore represents a memory-supported non-Condon tunneling pathway that remains active even when the bare electronic coupling is extremely weak.

When the tunneling element is increased to $\Delta=10^{-3}\,\mathrm{eV}$ in Fig.~\ref{fig:4}(b), both coupling geometries produce observable transfer. The diagonal interaction now operates together with the finite Hamiltonian tunneling term, leading to a gradual decrease of the donor population and corresponding growth of the acceptor population. Off-diagonal coupling produces a faster and somewhat larger transfer, superimposed with oscillations that follow the recurrence of the structured bath. The separation between the two cases is smaller than in Fig.~\ref{fig:4}(a), but the oscillatory modulation of the $L_x$ trajectory shows that the environmental coordinate continues to act as a dynamical tunneling gate.

At the intermediate tunneling strength, $\Delta=10^{-2}\,\mathrm{eV}$, Fig.~\ref{fig:4}(c) exhibits large-amplitude, non-exponential donor--acceptor exchange. Diagonal coupling permits nearly complete population transfer followed by pronounced back-transfer, demonstrating that the system retains substantial coherent and vibronic character. The off-diagonal interaction also drives a large initial transfer, but its subsequent oscillations are damped, shifted in phase, and reduced in amplitude. Thus, structured memory does not simply increase the acceptor population at every time. Instead, it can reinforce one part of the transfer cycle while suppressing or redirecting another, depending on the phase with which the environmental response returns.

In the strong-tunneling regime, $\Delta=10^{-1}\,\mathrm{eV}$, shown in Fig.~\ref{fig:4}(d), the intrinsic Hamiltonian produces rapid population oscillations for both coupling operators. The diagonal case retains comparatively large-amplitude exchange, whereas off-diagonal coupling produces stronger damping and drives the populations toward a more mixed distribution. The two trajectories also develop different phases because the structured bath acts through different system operators. In this regime, the bath no longer creates the primary transfer pathway; it mainly controls the coherence, phase, and damping of an exchange already generated by the strong electronic coupling.

Figure~\ref{fig:4} therefore demonstrates that the effect of non-Markovian memory depends on both the tunneling regime and the coupling geometry. Off-diagonal coupling provides its clearest transfer enhancement when the bare electronic pathway is weak. At intermediate and strong tunneling, the same memory-bearing interaction can instead produce damping, phase shifts, and back-transfer. Non-Markovianity should consequently be regarded as a mechanism for dynamically reshaping ligand--receptor electron transfer rather than as a universal source of rate enhancement.

These results show that non-Markovianity is a dynamical control variable, not a synonym for enhancement. Its effect is determined by the overlap of the bath correlation time with the tunneling and ligand-vibration timescales and by the operator through which the returning fluctuation acts. In the diagonal geometry, memory feeds back mainly through the bare tunneling commutator in Eq.~\eqref{eq:obar-z}. In the off-diagonal geometry, Eq.~\eqref{eq:obar-x} couples the feedback directly to the bias and ligand coordinate, making the response more sensitive to vibrational motion.

\subsection{Non-Condon coupling acts as a tunneling gate}

\begin{figure}[p]
\centering
\includegraphics[width=\linewidth]{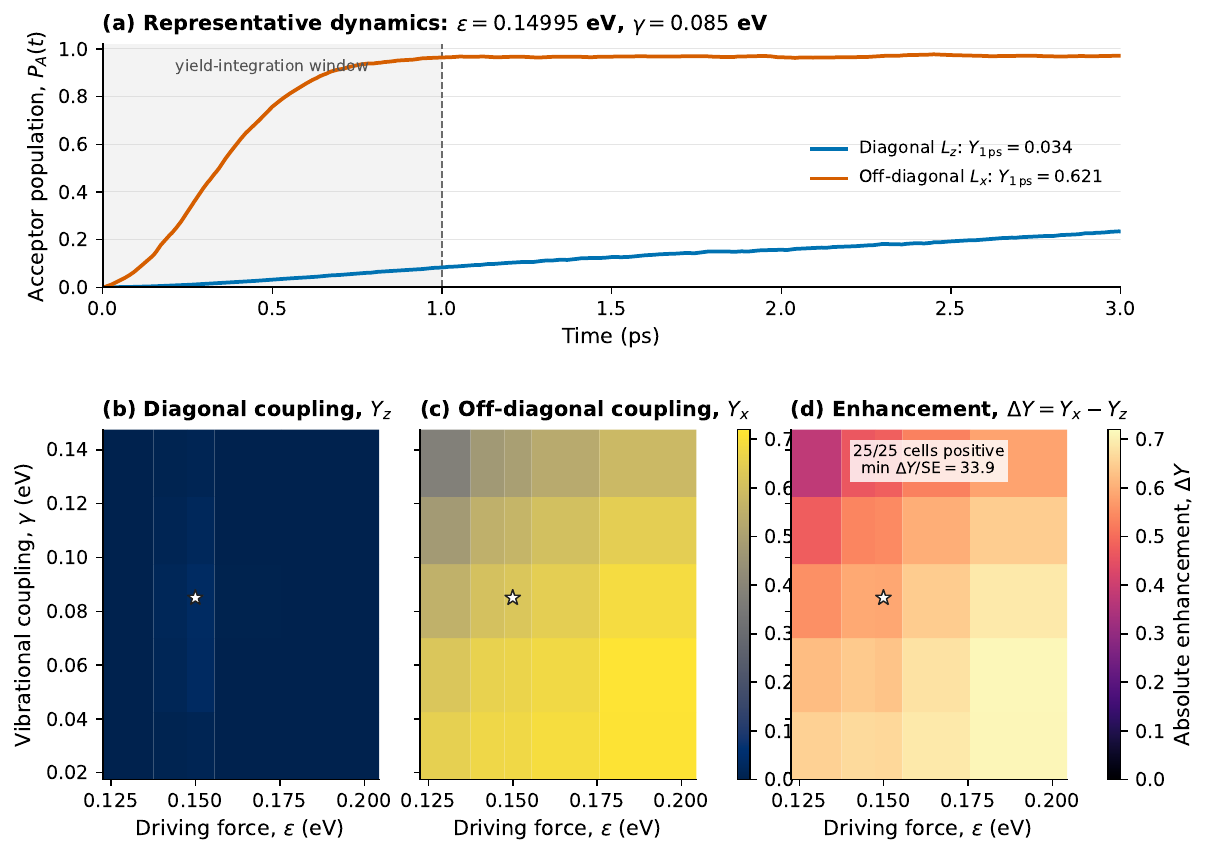}
\caption{\textbf{Finite-time ET enhancement from non-Condon coupling.} (a) Representative acceptor-population traces for $L_z$ and $L_x$. (b,c) Matched one-picosecond transfer-yield maps for diagonal and off-diagonal environmental coupling. (d) The direct enhancement $\Delta Y=Y_x-Y_z$, which is positive throughout the grid. The star identifies the traces in panel (a); parameters and paired uncertainties are given in Tables~S1 and S3 and Fig.~S5.}
\label{fig:5}
\end{figure}

To compare the diagonal and off-diagonal mechanisms directly, Fig.~\ref{fig:5} uses identical electronic, vibrational, bath, propagation, and sampling parameters for both coupling geometries. Because the resulting off-diagonal trajectories are not generally described by a single exponential, the comparison is based on the finite-time transfer yield defined in Eq.~\eqref{eq:yield}. This observable measures the acceptor population accumulated over the first picosecond without assuming irreversible kinetic relaxation.

Figure~\ref{fig:5}(a) shows the acceptor-population dynamics at the representative parameter point marked by the stars in the remaining panels. The shaded region identifies the interval used to calculate the one-picosecond time-averaged yield. Under diagonal coupling, the acceptor population grows slowly because $L_z$ changes the activation energy but relies on the weak bare tunneling element to connect the donor and acceptor. In contrast, the off-diagonal interaction produces a rapid increase in acceptor population because $L_x$ directly modulates the electronic-transition matrix element. The off-diagonal trajectory approaches a large acceptor population within the integration window, whereas the diagonal trajectory remains close to the donor state. Consequently,
\begin{equation}
Y_{1\mathrm{ps}}^{(z)}=0.034,
\qquad
Y_{1\mathrm{ps}}^{(x)}=0.621,
\end{equation}
corresponding to an approximately eighteen-fold enhancement of the finite-time transfer yield at this point.

Figure~\ref{fig:5}(b) presents the diagonal-coupling yield $Y_z$ as a function of the driving force $\epsilon$ and ligand-mode coupling $\gamma$. Appreciable transfer is confined to a narrow region near the one-quantum vibronic condition $\epsilon\simeq\lambda_s+\omega_v$. This localization occurs because diagonal fluctuations can bring the donor and acceptor energies into resonance but cannot themselves change the electronic state. The yield also varies non-monotonically with $\gamma$ because the ligand displacement controls the one-quantum Franck--Condon weight. Weak displacement provides insufficient vibronic overlap, whereas strong displacement redistributes intensity into higher-order vibrational channels. The diagonal mechanism is therefore both energy selective and sensitive to the ligand-mode displacement.

Figure~\ref{fig:5}(c) shows the corresponding off-diagonal yield $Y_x$ using the same parameter grid and color scale. In contrast to the narrow diagonal feature, the off-diagonal map displays substantial transfer throughout the investigated region. Nuclear fluctuations acting through $L_x$ directly open and close the tunneling pathway, so transfer is less restricted by the one-quantum activationless condition. The smooth variation of $Y_x$ with $\epsilon$ and $\gamma$ reflects the combined influence of electronic bias, vibronic dressing, and tunneling modulation rather than a single narrow resonance. The large difference between Figs.~\ref{fig:5}(b) and \ref{fig:5}(c) therefore arises from the operator through which the same environment acts on the electronic subsystem.

The direct comparison is shown in Fig.~\ref{fig:5}(d) through the absolute enhancement
\begin{equation}
\Delta Y_{1\mathrm{ps}}
=
Y_{1\mathrm{ps}}^{(x)}
-
Y_{1\mathrm{ps}}^{(z)}.
\end{equation}
Positive values occupy all $25$ matched parameter cells, demonstrating that off-diagonal coupling delivers more population to the acceptor during the observation window than diagonal coupling under the same conditions. The enhancement is particularly pronounced away from the narrow diagonal resonance, where energy-gap modulation is ineffective but tunneling modulation remains active. Paired trajectory-block analysis shows that even the least resolved enhancement remains well above its statistical uncertainty, as reported in the SI.

Figure~\ref{fig:5} therefore supports a non-Condon enhancement of finite-time electron-transfer efficiency, but it does not establish a universal enhancement of a kinetic rate constant. The off-diagonal dynamics can contain initial acceleration, saturation, oscillation, and back-transfer, so assigning a single exponential rate would discard the memory-dependent character of the motion. The physically supported conclusion is that fluctuations of the tunneling pathway transfer substantially more population to the acceptor over the relevant finite-time window than fluctuations of the energy gap. This behavior is the dynamical-gating mechanism implied by the coordinate dependence of $V_{DA}$ in Eq.~\eqref{eq:coordinate-expansion}.

\subsection{Limiting cases validate the propagated ingredients}

\begin{figure}[p]
\centering
\includegraphics[width=\linewidth]{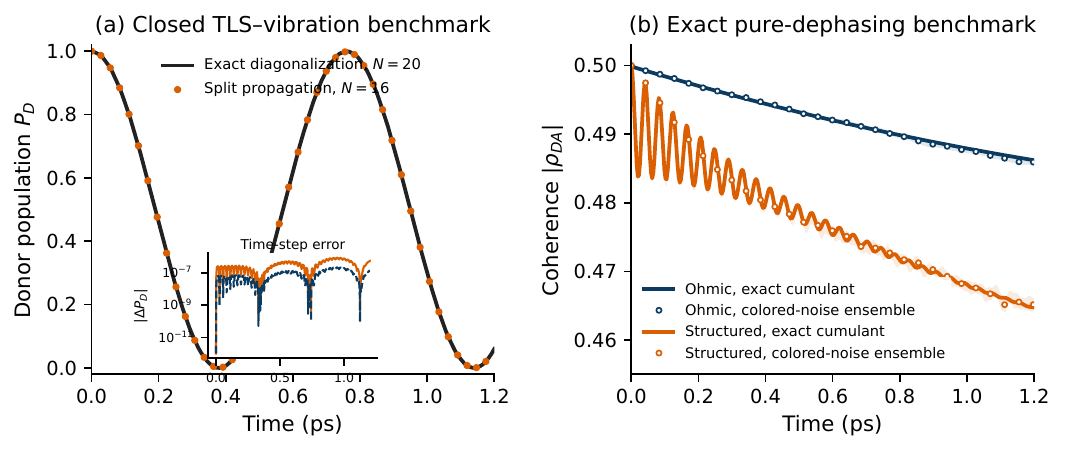}
\caption{\textbf{Exact limiting-case tests.} (a) Closed electronic--vibrational dynamics from split propagation and direct Hamiltonian evolution. (b) Pure-dephasing coherence from colored-noise sampling and the exact Gaussian cumulant for the Ohmic and structured baths. Insets and shaded bands show numerical errors and block uncertainties; values are tabulated in the SI.}
\label{fig:6}
\end{figure}

Figure~\ref{fig:6} examines two exactly solvable limits that isolate the deterministic and stochastic components of the numerical implementation. These tests are deliberately separated from the full electron-transfer problem so that agreement can be attributed to a specific part of the calculation rather than to cancellation among several approximations.

Figure~\ref{fig:6}(a) tests the deterministic propagation of the coupled two-level system and explicit ligand vibration after the residual bath is removed. The solid curve is obtained by direct diagonalization of the complete electronic--vibrational Hamiltonian using a larger vibrational basis, whereas the symbols are produced by the split-propagation algorithm used in the stochastic calculations. The donor population undergoes nearly complete oscillatory transfer to the acceptor and subsequent return to the donor. Agreement is maintained through the maxima, minima, and turning points of this motion, demonstrating that the numerical scheme correctly propagates the electronic tunneling, vibronic displacement, and their coupled dynamics over the physical time axis.

The inset of Fig.~\ref{fig:6}(a) shows the absolute population error relative to exact evolution in the same numerical basis. Reducing the time step decreases the error by approximately the factor expected for the second-order split scheme. Comparison with the larger-basis reference also confirms that truncating the explicit ligand mode does not materially alter the plotted population dynamics. This panel therefore tests the Hamiltonian construction, operator splitting, vibrational-basis convergence, initial-state preparation, and conversion between propagation units and picoseconds without introducing stochastic sampling or a memory closure.

Figure~\ref{fig:6}(b) provides an independent test of the environmental contribution. The tunneling and explicit vibronic coupling are removed, and the electronic system is initialized in a coherent donor--acceptor superposition. With $L_z=\sigma_z/2$, the interaction commutes with the system Hamiltonian, so the coherence is known exactly from the Gaussian cumulant,
\begin{equation}
\left|\rho_{DA}(t)\right|
=
\frac{1}{2}\exp[-\Phi(t)],
\end{equation}
where $\Phi(t)$ is determined by the same spectral density and temperature used to generate the colored stochastic processes. The solid curves show this analytical result, while the open symbols represent the colored-noise ensemble averages and the shaded regions indicate their sampling uncertainties.

For the Ohmic spectrum, the coherence displays a smooth decay because the bath correlation disappears rapidly. The structured spectrum produces an oscillatory modulation superimposed on the decay, reflecting the repeated phase-sensitive action of the narrow environmental mode. In both cases, the stochastic results follow the corresponding exact cumulant over the full propagation interval. Panel~\ref{fig:6}(b) therefore validates the frequency-domain noise synthesis, finite-temperature fluctuation amplitudes, sampled covariance, ensemble averaging, and time conversion for both short-memory and structured environments.

The two panels establish that the deterministic electronic--vibrational propagator and the colored-noise construction reproduce their respective exact limits. They do not, however, make the full noncommuting population calculation exact. When $[H_{\mathrm S},L]\neq0$, the NMQSD functional derivative is replaced by the first-order finite-memory operator in Eq.~\eqref{eq:o-expansion}, and that approximation is not tested by either limiting case in Fig.~\ref{fig:6}. The associated numerical errors, convergence checks, and unsuccessful hierarchy-convergence audit are therefore reported separately in Table~S3 and Fig.~S6 rather than being presented as an exact benchmark of the complete electron-transfer dynamics.

\section{Discussion}

The calculations provide a unified mechanistic description of vibrationally assisted electron transfer at an effective ligand--receptor interface. The explicit ligand mode displaces the donor and acceptor potential-energy surfaces and generates a sequence of vibronic transfer channels. Diagonal environmental coupling acts primarily on the activation coordinate by modulating the donor--acceptor energy gap; consequently, transfer remains strongly dependent on vibronic energy matching and Franck--Condon overlap. Off-diagonal non-Condon coupling instead modulates the electronic overlap or tunneling barrier and can therefore open a direct transition pathway even when the bare tunneling element is weak. Environmental memory determines whether these perturbations decay locally or return to the electronic--vibrational subsystem as delayed, phase-sensitive feedback.

This framework explains the phase-dependent and non-monotonic influence of the structured environment. When a returning fluctuation has a favorable phase and acts through $L_x$, it can reinforce ligand-assisted donor--acceptor mixing and increase population delivery to the acceptor. The same correlation can subsequently promote back-transfer, shift the phase of the population exchange, or damp coherent motion that would otherwise be sustained by the system Hamiltonian. A structured spectrum and non-Condon coupling are therefore not, by themselves, sufficient conditions for an increased long-time chemical rate. The robust result of the matched comparison is instead an enhancement of acceptor occupation over a specified finite observation window.

The distinction between a kinetic rate and a finite-time transfer yield is essential when interpreting ligand--receptor dynamics. A biochemical rate constant presupposes that short-lived transients can be separated from a subsequent exponential relaxation. Long-lived environmental correlations, coherent oscillations, and partial reversibility violate this separation and make an exponential fit dependent on the selected fitting interval. The observable introduced in Eq.~\eqref{eq:yield} avoids this ambiguity by defining a reproducible functional of the acceptor population,

\begin{equation*}
Y_\tau=\frac{1}{\tau}\int_0^\tau P_A(t)\,dt,
\end{equation*}

which remains meaningful for monotonic, oscillatory, and partially reversible dynamics. This quantity is a time-averaged electronic occupation rather than an irreversible chemical or quantum yield. Relating it to downstream biochemical activity would require an explicit trapping, charge-separation, or reaction step.

The biological interpretation of the model is therefore strongest at the level of molecular design principles. A ligand vibration can promote diagonal vibronic transfer when its quantum energy satisfies the donor--acceptor energy requirement and its displacement produces appreciable Franck--Condon weight. Non-Condon enhancement additionally requires a vibrational or collective coordinate with a substantial electronic-coupling derivative, $\partial V_{DA}/\partial q$. Environmental memory becomes dynamically relevant when the corresponding correlation persists on a timescale comparable to the electronic tunneling or ligand-vibrational motion. These quantities can, in principle, be estimated using constrained density-functional theory, fragment-orbital calculations, normal-mode analysis, and molecular-dynamics trajectories.

Several limitations define the scope of the present conclusions. The spectral densities are analytical rather than atomistically derived, and the Ohmic and structured environments differ in both spectral shape and integrated coupling strength; their comparison therefore identifies contrasting dynamical regimes rather than isolating correlation time as a single variable. The diagonal and off-diagonal interactions are treated separately, cross-correlations are neglected, the ligand is represented by one harmonic mode, and the noncommuting population dynamics employ a first-order finite-memory closure. A quantitative application to a specific receptor should determine the matrix spectral densities $J_{zz}(\omega)$, $J_{xx}(\omega)$, and $J_{zx}(\omega)$ from molecular trajectories, incorporate relevant multiple-mode and anharmonic effects, introduce any irreversible chemical sink explicitly, and benchmark representative parameter points against a converged hierarchy or path-integral calculation. The present work supplies the operator derivation, physically appropriate observables, and numerical validation framework required for such a system-specific extension.

\section{Conclusion}

This study establishes a quantum-dynamical framework for examining vibrationally assisted electron transfer in an effective ligand--receptor complex. The receptor electron-transfer unit is represented by donor and acceptor diabatic states, a bound-ligand vibration is treated explicitly, and the surrounding protein--solvent environment is incorporated through a finite-temperature colored memory kernel. Linearization of the coordinate-dependent diabatic Hamiltonian provides a direct physical distinction between diagonal coupling, which modulates the donor--acceptor energy gap, and off-diagonal non-Condon coupling, which modulates the electronic tunneling matrix element.

In the weak-tunneling, rapidly equilibrating diagonal-bath limit, the model recovers the established Marcus--Levich--Jortner description of vibronic transfer. Persistent environmental correlations produce qualitatively different behavior, including transient oscillations, back-transfer, phase shifts, and non-exponential relaxation. These results demonstrate that non-Markovian memory is not intrinsically beneficial or detrimental to electron transfer. Its influence is determined jointly by the bath correlation time, the ligand-vibrational timescale, the bare electronic coupling, and the operator through which the environment acts. Memory can support a weak transfer pathway, but it can also damp or redirect coherent donor--acceptor exchange.

The most pronounced enhancement occurs when the bare tunneling interaction is weak and the environment couples through the off-diagonal operator. In this regime, non-Condon fluctuations directly activate the electronic transition and generate a substantially larger finite-time acceptor population than diagonal energy-gap fluctuations under matched conditions. As the bare tunneling strength increases, intrinsic Hamiltonian mixing becomes dominant, and the principal effect of the environment shifts from opening a transfer pathway to controlling its phase, damping, and coherence. Because the off-diagonal dynamics are generally non-exponential, this enhancement is appropriately expressed as an increased finite-time transfer yield rather than as a universal increase in a single electron-transfer rate constant.

Exact closed-system and commuting pure-dephasing limits validate the deterministic propagator and colored-noise construction, respectively. The full noncommuting dynamics nevertheless retain the limitations of the first-order finite-memory closure, analytical spectral densities, and single effective ligand mode. The present model should therefore be interpreted as a mechanistic description rather than a quantitative simulation of a specific receptor. A system-specific extension will require molecular-dynamics and electronic-structure calculations to determine the relevant ligand modes, diabatic-state parameters, and diagonal, off-diagonal, and cross-correlated spectral densities, followed by validation against a converged hierarchy or path-integral method. Within these boundaries, the results identify non-Condon tunneling modulation as a plausible dynamical gate through which ligand vibration and environmental memory can cooperate to regulate electron transfer in receptor-like molecular environments.

\section*{Data and code availability}

All numerical source data, convergence diagnostics, and executable Python scripts used to generate the figures are included in the accompanying repository archive. The datasets and parameter values utilized in this study are used from previously published sources as cited. Additional information is available at \url{https://github.com/waqashaseeb}.

\section*{Author contributions}

M.W.H.: conceptualization, methodology, software, validation, formal analysis, investigation, data curation, visualization, and writing--original draft. M.T.: conceptualization, methodology, supervision, project administration, and writing--review and editing. Both authors approved the submitted version.

\section*{Acknowledgments}

This work was funded by United Arab Emirates University Research Affairs under Grant G-00003550.

\section*{Declaration of interests}

The authors declare no competing interests.

\bibliographystyle{unsrtnat}
\bibliography{references}

\end{document}


\maketitle

\section*{Contents}

This Supporting Information provides the full ligand--receptor coupling derivation, displaced-oscillator convention, finite-temperature colored-noise equation, diagonal and off-diagonal memory operators, rate-analysis rules, physical and numerical parameters, validation metrics, and five supplementary figures. Machine-readable source data and scripts are included in the accompanying repository-ready archive.

\section*{Note S1. From a ligand--receptor diabatic model to diagonal and off-diagonal coupling}

For two localized electronic configurations, the nuclear-coordinate-dependent Hamiltonian is
\begin{equation}
H_{\mathrm{el}}(\mathbf q)=
\frac{E_D(\mathbf q)+E_A(\mathbf q)}{2}I
+\frac{E_D(\mathbf q)-E_A(\mathbf q)}{2}\sigma_z
+V_{DA}(\mathbf q)\sigma_x.
\end{equation}
The first term contributes only a common nuclear potential and is omitted from the electronic dynamics. Around a reference ligand--receptor geometry $\mathbf q_0$, define $\delta q_k=q_k-q_{0k}$. The energy gap and twice the electronic coupling have the linear expansions
\begin{align}
E_D(\mathbf q)-E_A(\mathbf q)
&\simeq \epsilon+\sum_k
\left.\frac{\partial(E_D-E_A)}{\partial q_k}\right|_{\mathbf q_0}\delta q_k,\\
2V_{DA}(\mathbf q)
&\simeq \Delta+\sum_k
\left.\frac{\partial(2V_{DA})}{\partial q_k}\right|_{\mathbf q_0}\delta q_k.
\end{align}
After writing $\delta q_k=q_{\mathrm{zpf},k}(B_k+B_k^\dagger)$, the coupling constants are
\begin{equation}
c_{zk}=q_{\mathrm{zpf},k}
\left.\frac{\partial(E_D-E_A)}{\partial q_k}\right|_{\mathbf q_0},
\qquad
c_{xk}=q_{\mathrm{zpf},k}
\left.\frac{\partial(2V_{DA})}{\partial q_k}\right|_{\mathbf q_0}.
\end{equation}
The interaction is therefore
\begin{equation}
H_{\mathrm{SB}}=\frac{\sigma_z}{2}\sum_kc_{zk}(B_k+B_k^\dagger)
+\frac{\sigma_x}{2}\sum_kc_{xk}(B_k+B_k^\dagger).
\label{eq:si-general-coupling}
\end{equation}
The first term is diagonal in the localized donor--acceptor basis and changes the energy gap. The Condon approximation neglects the coordinate derivative of $V_{DA}$; retaining that derivative produces the second, off-diagonal non-Condon term. In a fully parameterized molecular model both terms can be present, with matrix spectral densities
\begin{equation}
J_{\alpha\beta}(\omega)=\pi\sum_kc_{\alpha k}c_{\beta k}
\delta(\omega-\omega_k),
\qquad \alpha,\beta\in\{z,x\}.
\end{equation}
The present calculations set either $c_{zk}$ or $c_{xk}$ to zero so that the two limiting mechanisms can be compared without cross-correlation.

\section*{Note S2. Displaced ligand oscillator and Huang--Rhys convention}

For a fixed eigenvalue $s=\pm1$ of $\sigma_z$, the explicit ligand-mode Hamiltonian is
\begin{equation}
H_v^{(s)}=\omega_v b^\dagger b+s\gamma(b+b^\dagger)
=\omega_v\left(b^\dagger+s\frac{\gamma}{\omega_v}\right)
\left(b+s\frac{\gamma}{\omega_v}\right)-\frac{\gamma^2}{\omega_v}.
\end{equation}
The two minima are displaced by $2d$, where $d=\gamma/\omega_v$. Their vibrational overlaps are
\begin{equation}
\left|\langle m;+d|-d\rangle\right|^2=e^{-S}\frac{S^m}{m!},
\qquad
S=(2d)^2=\left(\frac{2\gamma}{\omega_v}\right)^2.
\end{equation}
With this symmetric $\pm\gamma$ convention, the explicit-mode reorganization energy is
\begin{equation}
\lambda_v=S\omega_v=\frac{4\gamma^2}{\omega_v}.
\end{equation}
This factor follows from the separation of the two minima and differs from conventions in which only one electronic surface is displaced. The donor-equilibrated initial state is $\ket D\otimes\ket{-d}$. In the controlled MLJ benchmark, the center of vibronic channel $m$ is $\epsilon_m=\lambda_s+m\omega_v$; changing $\gamma$ changes the Franck--Condon weight rather than the channel center.

\section*{Note S3. Spectral densities, thermal correlations, and memory times}

The spectral-density normalization is
\begin{equation}
J(\omega)=\pi\sum_kc_k^2\delta(\omega-\omega_k),
\qquad
\lambda=\frac{1}{\pi}\int_0^\infty\frac{J(\omega)}{\omega}\,d\omega.
\end{equation}
At temperature $T$, the bath correlation is decomposed into zero-point and thermal parts,
\begin{align}
C(t)&=C_0(t)+C_T(t),\\
C_0(t)&=\frac{1}{\pi}\int_0^\infty J(\omega)e^{-i\omega t}\,d\omega,\\
C_T(t)&=\frac{2}{\pi}\int_0^\infty J(\omega)n_{\mathrm B}(\omega)
\cos(\omega t)\,d\omega,
\end{align}
where $n_{\mathrm B}(\omega)=[\exp(\omega/k_{\mathrm B}T)-1]^{-1}$. Their sum is equivalent to the usual expression containing $\coth(\omega/2k_{\mathrm B}T)$.

For the exponential-cutoff Ohmic form, the characteristic short-memory scale is $\tau_{\mathrm{Ohm}}\sim\hbar/\omega_c$. For the underdamped Brownian oscillator, the approximate oscillation period and envelope time are
\begin{equation}
\tau_{\mathrm{osc}}=\frac{2\pi\hbar}{\Omega_0},
\qquad
\tau_{\mathrm{env}}\simeq\frac{2\hbar}{\Gamma}.
\end{equation}
For the Figure~2 parameters these are $1.32\times10^{-3}$ ps, $0.0413$ ps, and $0.263$ ps, respectively. Direct quadrature reproduces the specified reorganization energies with relative errors $1.29\times10^{-7}$ for the Ohmic spectrum and $2.11\times10^{-5}$ for the structured spectrum.

\section*{Note S4. Finite-temperature nonlinear colored-noise propagation}

A circular complex Gaussian process $z_t$ represents $C_0$ and an independent real Gaussian process $\eta_t$ represents $C_T$:
\begin{align}
\mathbb E[z_tz_s]&=0,& \mathbb E[z_tz_s^*]&=C_0(t-s),\\
\mathbb E[\eta_t]&=0,& \mathbb E[\eta_t\eta_s]&=C_T(t-s).
\end{align}
Both processes are synthesized in the frequency domain from nonnegative discrete spectral weights. Antithetic paths use opposite Gaussian amplitudes and therefore preserve the target covariance.

For Hermitian $L$, define $\Delta_t L=L-\langle L\rangle_t$ and the shifted zero-point noise
\begin{equation}
\widetilde z_t^*=z_t^*+\int_0^t C_0^*(t-s)\langle L\rangle_s\,ds.
\end{equation}
The normalized first-order finite-memory equation implemented for the noncommuting calculations is
\begin{equation}
\frac{d}{dt}\ket{\widetilde\psi_t}=
\left[-iH_{\mathrm S}-iL\eta_t
+\Delta_tL\,\widetilde z_t^*
-\Delta_tL\,\overline O^{(1)}(t)
+\left\langle\Delta_tL\,\overline O^{(1)}(t)\right\rangle_t
\right]\ket{\widetilde\psi_t},
\label{eq:si-nonlinear}
\end{equation}
with
\begin{equation}
\overline O^{(1)}(t)=g_0(t)L-i g_1(t)[H_{\mathrm S},L],
\qquad
g_n(t)=\int_0^t\tau^nC_0(\tau)\,d\tau.
\end{equation}
The thermal process appears as a Hermitian stochastic potential. The memory shift in $\widetilde z_t^*$ is propagated through an exponential representation of $C_0(t)$.

For $L_z=\sigma_z/2$, only the electronic tunneling term fails to commute with $L_z$:
\begin{align}
[H_{\mathrm S},L_z]
&=\frac{\Delta}{4}[\sigma_x,\sigma_z]
=-i\frac{\Delta}{2}\sigma_y,\\
\overline O_z^{(1)}(t)
&=\frac{g_0(t)}{2}\sigma_z-\frac{\Delta g_1(t)}{2}\sigma_y.
\label{eq:si-obar-z}
\end{align}
For $L_x=\sigma_x/2$, the bias and ligand-mode terms contribute,
\begin{align}
[H_{\mathrm S},L_x]
&=\frac{\epsilon}{4}[\sigma_z,\sigma_x]
+\frac{\gamma}{2}[\sigma_z,\sigma_x](b+b^\dagger)\\
&=i\left[\frac{\epsilon}{2}\sigma_y
+\gamma\sigma_y(b+b^\dagger)\right],\\
\overline O_x^{(1)}(t)
&=\frac{g_0(t)}{2}\sigma_x
+g_1(t)\left[\frac{\epsilon}{2}\sigma_y
+\gamma\sigma_y(b+b^\dagger)\right].
\label{eq:si-obar-x}
\end{align}
Equations~\eqref{eq:si-obar-z} and \eqref{eq:si-obar-x} show that the first memory correction has different operator content in the two coupling geometries. When $C_0(\tau)$ becomes sharply localized, $g_1$ vanishes relative to $g_0$ and the correction becomes local. Equation~\eqref{eq:si-nonlinear} is a declared first-order closure of the exact NMQSD functional derivative; it is not a formally exact solution of the full noncommuting spin--boson problem.

\section*{Note S5. Controlled MLJ reference and rate extraction}

The MLJ comparison is performed in a separate high-temperature classical-solvent limit. The real energy-gap covariance is
\begin{equation}
\langle\xi(t)\xi(0)\rangle=
\frac{2\lambda_s k_{\mathrm B}T}{1+(\omega_ct)^2},
\end{equation}
and each path is propagated with
\begin{equation}
H_\xi(t)=H_{\mathrm S}+\frac{1}{2}[\xi(t)-\lambda_s]\sigma_z.
\end{equation}
The $-\lambda_s$ shift and the donor-displaced oscillator prepare the classical solvent and explicit mode consistently. Stationary paths are generated by circulant embedding.

For small donor depletion, the simulated forward rate is the least-squares slope
\begin{equation}
k_f=\frac{\sum_{t_n\in\mathcal W}(t_n-\bar t)P_A(t_n)}
{\sum_{t_n\in\mathcal W}(t_n-\bar t)^2},
\end{equation}
with $\mathcal W=0.20$--$1.00$ ps. Block standard errors are obtained from antithetic trajectory groups. The rate map and MLJ expression have Pearson correlation $r=0.9992$ and peak-normalized root-mean-square difference $1.37\%$. The empirical classical covariance differs from its target by $0.179\%$, and increasing the explicit-mode basis at the most demanding tested point changes the rate by $0.49\%$.

\section*{Note S6. Rate admissibility, transfer yield, and exact limits}

Outside the controlled MLJ calculation, a scalar rate is accepted only when every criterion in Table~\ref{tab:criteria} is satisfied. When the maximum acceptor population over 3 ps is at most 0.10, a small-depletion slope is used. Otherwise, the population is fit to
\begin{equation}
P_A(t)=P_\infty+[P_A(t_0)-P_\infty]
e^{-k_{\mathrm{rel}}(t-t_0)},
\qquad
k_f=P_\infty k_{\mathrm{rel}}.
\end{equation}
The common-bath scan in Fig.~S4 produced no setting accepted for both coupling geometries. Candidate fitted values in rejected cells are retained only to diagnose the failure of a rate description.

The finite-time transfer observable is
\begin{equation}
Y_{1\mathrm{ps}}^{(\alpha)}=\frac{1}{1\ \mathrm{ps}}
\int_0^{1\ \mathrm{ps}}P_A^{(\alpha)}(t)\,dt.
\end{equation}
Diagonal and off-diagonal simulations use the same antithetic noise blocks. The enhancement $\Delta Y=Y^{(x)}-Y^{(z)}$ is evaluated within each of 16 paired blocks, preserving their covariance. All 25 cells have positive $\Delta Y$; the least resolved difference has $\Delta Y/\mathrm{SE}(\Delta Y)=33.9$. At the representative point used in the main figure, $Y_z=0.03450$ and $Y_x=0.62100$.

The commuting pure-dephasing limit uses $L_z=\sigma_z/2$ and the initial electronic state $(\ket D+\ket A)/\sqrt2$. Its exact coherence magnitude is
\begin{equation}
|\rho_{DA}(t)|=\frac12e^{-\Phi(t)},
\qquad
\Phi(t)=\int_0^\infty\frac{d\omega}{\pi}
\frac{J(\omega)}{\omega^2}
\coth\!\left(\frac{\omega}{2k_{\mathrm B}T}\right)
[1-\cos(\omega t)].
\end{equation}
This limit tests the colored processes without invoking noncommuting transfer dynamics.

\clearpage

\begin{table}[ht]
\centering
\caption{Figure-specific parameters. Energies are in eV; time steps are in eV$^{-1}$ before conversion with $\hbar=6.582119569\times10^{-4}$ eV ps.}
\label{tab:parameters}
\scriptsize
\begin{tabular}{p{0.10\linewidth}p{0.78\linewidth}}
\toprule
Figure & Parameters \\
\midrule
1 / S1 & $T=300$ K; $\Delta=0.001$; $\lambda_s=0.0125$; $\omega_c=0.005$; $\omega_v=0.1487$; $\epsilon=0.09$--0.24; $\gamma=0.02$--0.14; 8 mode states; $\Delta t=2$; 2048 trajectories per map cell and 4096 for the profile. \\
2 & Ohmic: $\lambda=1.25\times10^{-4}$, $\omega_c=0.50$. Structured: $\lambda=0.0016$, $\Omega_0=0.10$, $\Gamma=0.005$. $T=290$ K. \\
3 / S2 / S3 & $\epsilon=0.1487$; $\omega_v=0.1087$; $\gamma=0.10$; $\Delta=10^{-4},10^{-3},10^{-2},10^{-1}$; Ohmic bath as in Fig. 2; $T=290$ K; 16 mode states; $\Delta t=0.5$; 1024 trajectories; 1.20 ps. \\
4 / S2 / S3 & $\epsilon=\omega_v=0.1487$; $\gamma=0.10$; same $\Delta$ values; structured bath as in Fig. 2; $T=290$ K; 16 mode states; $\Delta t=0.5$; 1024 trajectories; 1.20 ps. \\
5 / S5 & $T=300$ K; $\Delta=0.001$; $\lambda_s=0.00125$; $\omega_c=0.05$; $\omega_v=0.1487$; $\epsilon=0.13,0.145,0.14995,0.1612,0.19$; $\gamma=0.03,0.055,0.085,0.11,0.135$; 10 mode states; $\Delta t=0.5$; 256 trajectories in 16 paired blocks; 3 ps. \\
6 & Closed system: $\epsilon=\omega_v=0.1487$, $\gamma=0.10$, $\Delta=0.01$. Pure dephasing: $\Delta=\gamma=0$, $T=290$ K, 4096 trajectories, with the Fig. 2 baths. \\
S4 & $\epsilon=0.14995$; $\omega_v=0.1487$; $\gamma=0.085$; $\Delta=0.001$; $T=300$ K; $\lambda_s=1.25\times10^{-4},1.25\times10^{-3},1.25\times10^{-2}$; $\omega_c=0.05,0.20,0.50$; 256 trajectories; 3 ps. \\
\bottomrule
\end{tabular}
\end{table}

\begin{table}[ht]
\centering
\caption{Every criterion must be satisfied before a fitted value is called an ET rate.}
\label{tab:criteria}
\small
\begin{tabular}{p{0.28\linewidth}p{0.30\linewidth}p{0.30\linewidth}}
\toprule
Criterion & Small-depletion branch & Finite-depletion branch \\
\midrule
Central window & 0.20--2.80 ps & 0.05--3.00 ps \\
Alternative windows & 0.15--2.50; 0.30--3.00 ps & 0.03--2.50; 0.10--3.00 ps \\
Signal & $k_f\geq2\,\mathrm{SE}(k_f)$ & $k_f\geq2\,\mathrm{SE}(k_f)$ \\
Goodness of fit & $R^2\geq0.985$ & $R^2\geq0.98$ and NRMSE $\leq0.12$ \\
Window stability & Relative variation $\leq0.25$ & Relative variation $\leq0.20$ \\
Monotonicity & At least 90\% & At least 90\% \\
Additional condition & $\max P_A\leq0.30$ in fit window & At least one relaxation time observed \\
\bottomrule
\end{tabular}
\end{table}

\begin{table}[ht]
\centering
\caption{Validation and convergence metrics supporting the main figures.}
\label{tab:validation}
\scriptsize
\begin{tabular}{p{0.19\linewidth}p{0.69\linewidth}}
\toprule
Calculation & Diagnostic \\
\midrule
MLJ rate map & Pearson $r=0.9992$; peak-normalized NRMSE $=1.37\%$; colored-covariance NRMSE $=0.179\%$; one-quantum ridge RMSE $=0$; 8-to-12-state rate change $=0.49\%$ at the most demanding tested point. \\
Figure 3 & Quantum/thermal covariance errors $2.25\%/0.59\%$; exponential-memory fit error $0.153\%$; halving $\Delta t$ changes $P_D$ by at most $7.0\times10^{-4}$; 16-to-20-state change below $8.0\times10^{-4}$ over 1.2 ps. \\
Figure 4 & Quantum/thermal covariance errors $0.80\%/2.63\%$; exponential-memory fit error $0.217\%$; halving $\Delta t$ changes $P_D$ by at most $1.4\times10^{-4}$; 16-to-20-state change below $2.9\times10^{-7}$ over 1.2 ps. \\
Figure 5 & All 25 paired yield differences are positive; minimum $\Delta Y/\mathrm{SE}=33.9$; maximum pre-normalization norm error $8.1\times10^{-7}$; memory-fit error $0.176\%$. \\
Figure 6 & Split propagator versus direct 20-state propagation: maximum population error $9.1\times10^{-7}$; pure-dephasing maximum coherence errors $5.96\times10^{-4}$ (Ohmic) and $1.37\times10^{-3}$ (structured). \\
Common-bath scan & Accepted settings: 0/9 for $L_z$, 0/9 for $L_x$, and 0/9 jointly. Candidate fitted slopes are diagnostics, not accepted rates. \\
\bottomrule
\end{tabular}
\end{table}

\clearpage
\begin{figure}[p]
\centering
\includegraphics[width=\linewidth]{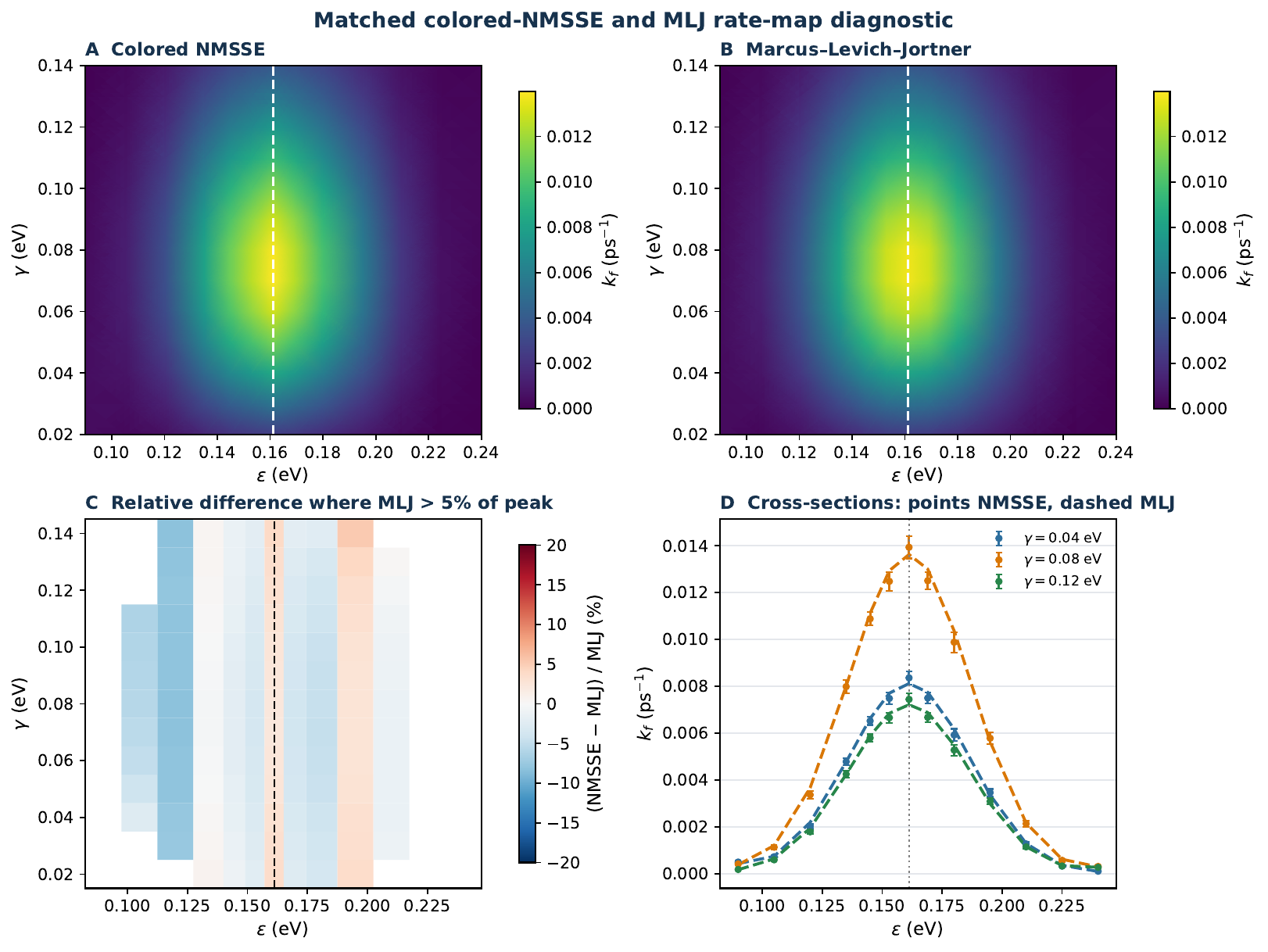}
\caption{\textbf{Detailed colored-solvent/MLJ comparison.} Simulated and theoretical rate maps are shown on common grids and scales with their difference, ridge positions, statistical consistency, and explicit-mode basis test.}
\label{fig:s1}
\end{figure}

\begin{figure}[p]
\centering
\includegraphics[width=\linewidth]{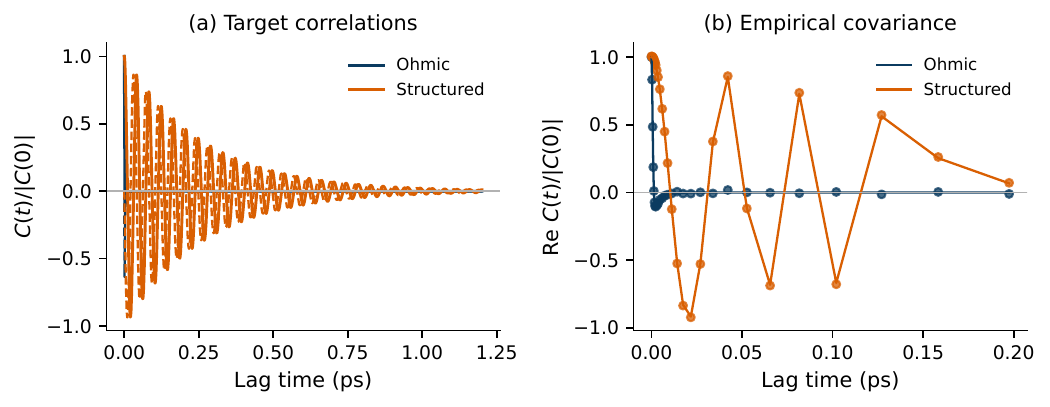}
\caption{\textbf{Colored-noise diagnostics for the Ohmic and structured population calculations.} Empirical quantum and thermal covariances are compared with their target functions at the lags used in the production checks.}
\label{fig:s2}
\end{figure}

\begin{figure}[p]
\centering
\includegraphics[width=\linewidth]{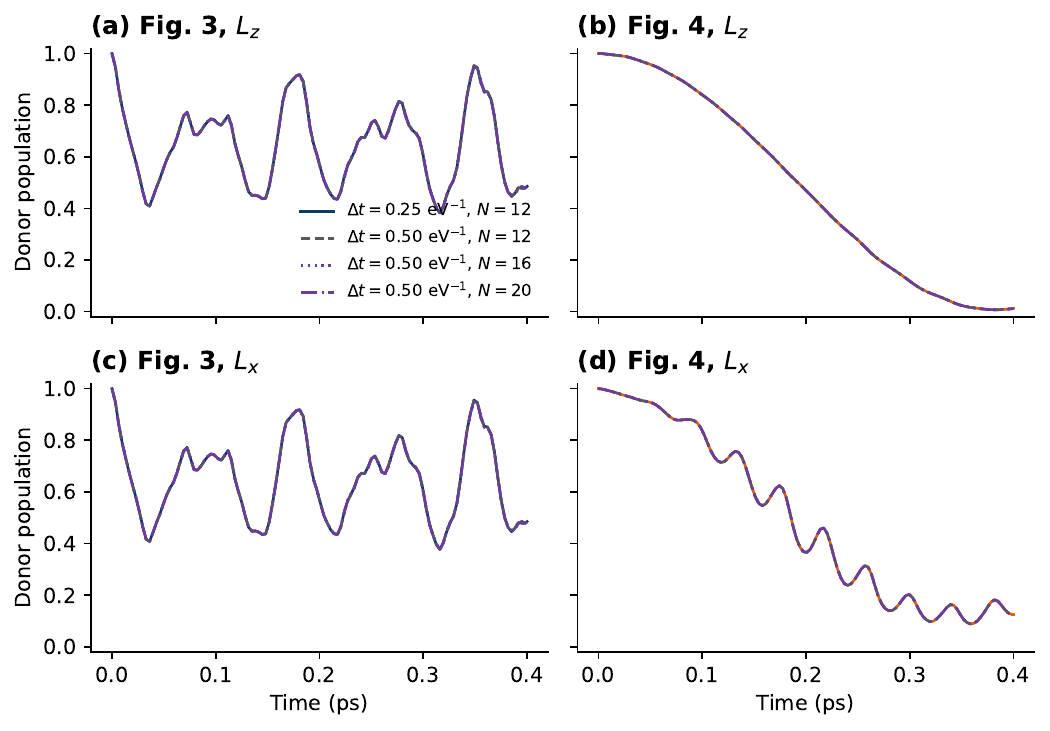}
\caption{\textbf{Time-step, ligand-mode basis, and trajectory-count checks.} The most demanding selected electronic-coupling case is tested for each bath; maximum differences are summarized in Table~\ref{tab:validation}.}
\label{fig:s3}
\end{figure}

\begin{figure}[p]
\centering
\includegraphics[width=\linewidth]{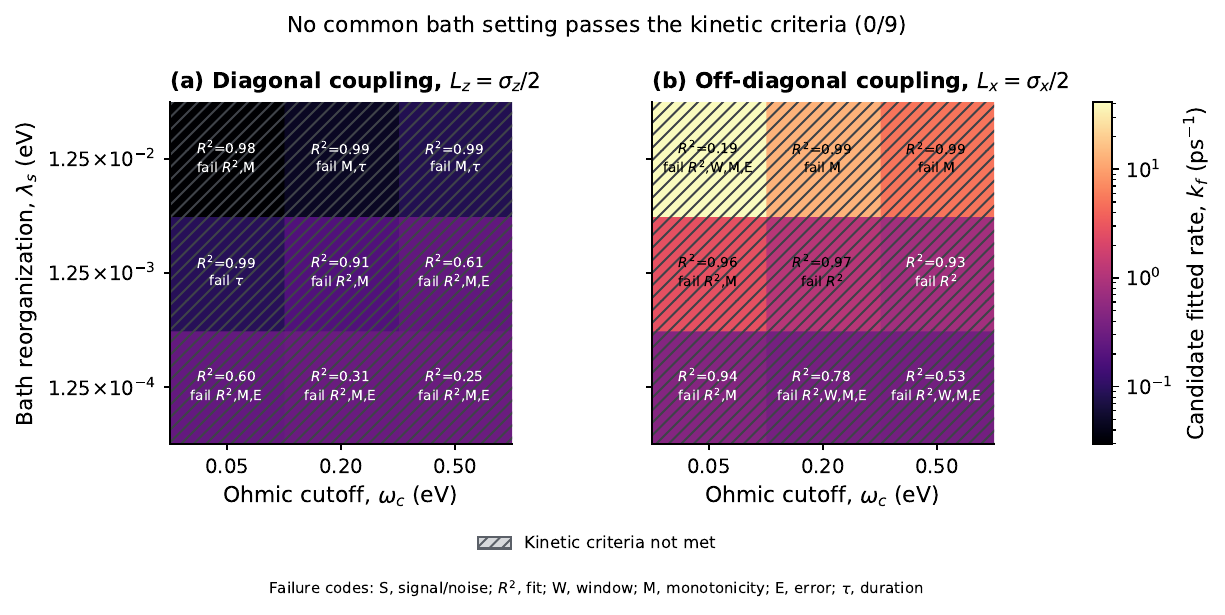}
\caption{\textbf{Common-bath kinetic-regime scan.} Diagonal and off-diagonal calculations use identical bath parameters. Hatched or crossed cells fail one or more criteria in Table~\ref{tab:criteria}. Fitted values in rejected cells are diagnostic and are not interpreted as chemical ET rates.}
\label{fig:s4}
\end{figure}

\begin{figure}[p]
\centering
\includegraphics[width=\linewidth]{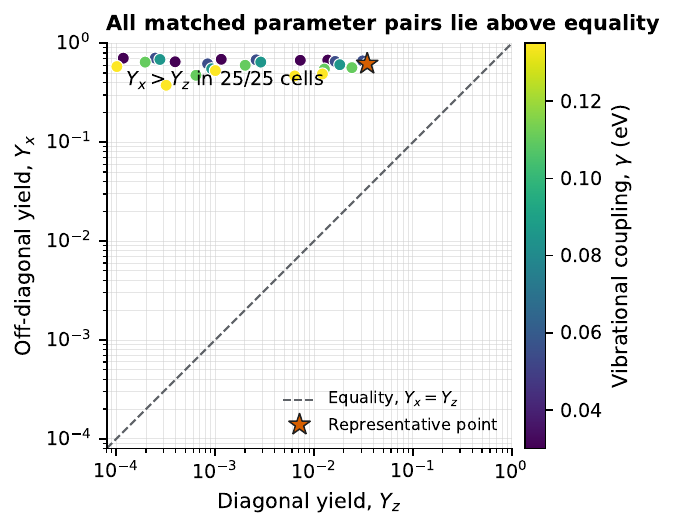}
\caption{\textbf{Paired comparison of the one-picosecond transfer yield.} Each point compares diagonal and off-diagonal coupling at the same electronic and ligand-mode parameters, using paired colored-noise blocks. Every point lies above the equality line.}
\label{fig:s5}
\end{figure}
